# Fractal dimension evolution and spatial replacement dynamics of urban growth

Yanguang Chen

(Department of Geography, College of Urban and Environmental Sciences, Peking University, Beijing 100871, P.R.China. E-mail: chenyg@pku.edu.cn)

**Abstract**: This paper presents a new perspective of looking at the relation between fractals and chaos by means of cities. Especially, a principle of space filling and spatial replacement is proposed to explain the fractal dimension of urban form. The fractal dimension evolution of urban growth can be empirically modeled with Boltzmann's equation. For the normalized data, Boltzmann's equation is equivalent to the logistic function. The logistic equation can be transformed into the well-known 1-dimensional logistic map, which is based on a 2-dimensional map suggesting spatial replacement dynamics of city development. The 2-dimensional recurrence relations can be employed to generate the nonlinear dynamical behaviors such as bifurcation and chaos. A discovery is made that, for the fractal dimension growth following the logistic curve, the normalized dimension value is the ratio of space filling. If the rate of spatial replacement (urban growth) is too high, the periodic oscillations and chaos will arise, and the city system will fall into disorder. The spatial replacement dynamics can be extended to general replacement dynamics, and bifurcation and chaos seem to be related with some kind of replacement process.

## 1. Introduction

Chaos and fractal geometry are always mentioned in the same literature because the chaotic attractors have fractal properties. However, many studies suggest that the concurrences of chaotic processes and fractal patterns do not imply the certain relation between chaos and fractals. As Bak (1996, page 31) pointed out: "In the popular literature, one finds the subjects of chaos and fractal geometry linked together again and again, despite the fact that they have little to do with each other." In fact, some chaos possesses no strange attractor (Chen, 2009). Notwithstanding this, it is necessary to detect the possible link between fractal structure and chaotic dynamics.

Cities are typical examples of random fractals, which can be described by various fractal parameters (Batty, 1995; Batty, 2008; Batty and Longley, 1994; Chen, 2010; Chen and Lin, 2009; De Keersmaecker *et al*, 2003; Frankhauser, 1994; Frankhauser, 1998; Terzi and Kaya, 2011; Thomas *et al*, 2007; Thomas *et al*, 2008). The significance of fractal city research lies in two aspects. On the one hand, the fractal structure suggests spatial optimization of nature. A fractal body can occupy its space in the most efficient way. Fractal city theory is helpful for future city planning. On the other hand, cities are complex spatial systems familiar to scientists. A city is a good window for us to recognize non-linear dynamics (Dendrinos and El Naschie, 1994). Chaos and fractals can be linked by cities and in turn provides novel ideas for us to understand cities.

Fractal patterns used to be revealed from chaotic process, but this time, I derive chaotic dynamics from a fractal dimension model. This paper will illustrate the following process. First, the fractal dimension growth of urban form can be mathematically described by Boltzmann's equation. For the normalized data, Boltzmann's equation can be transformed into a logistic function. Second, Discretizing the logistic function gives a 1-dimensional map, which can be deduced from a 2-dimenaional map associated with space-filling process. Both the 1-dimensional map and 2-dimensional map can yield complex patterns such as bifurcation and chaos. Third, the fractal dimension of urban form can be reinterpreted with spatial replacement dynamics, and the nonlinear dynamics show the relationship between fractal dimension and chaotic process.

## 2. Models

### 2.1 Boltzmann's equation of fractal dimension evolution

In theory, the box dimension of urban form ranges from 0 to 2. However, in practice, the box dimension always comes between 1 and 2. Boltzmann's equation is a possible choice for describing the fractal dimension growth of cities (Appendix 1). Boltzmann's equation was employed to model urban population evolution by Benguigui *et al* (2001). Urban population is associated with urban form. The Boltzmann model of fractal dimension evolution is as follows

$$D(t) = D_{\min} + \frac{D_{\max} - D_{\min}}{1+[\frac{D_{\max} - D_{(0)}}{D_{(0)} - D_{\min}}]e^{-kt}} = D_{\min} + \frac{D_{\max} - D_{\min}}{1+\exp(-\frac{t-t_0}{p})}, \quad (1)$$

where $D(t)$ refers to the fractal dimension of urban form in time of $t$, $D_{(0)}$ to the fractal dimension in the initial time/year, $D_{\max} \leq 2$ to upper limit of fractal dimension, i.e. the capacity of fractal dimension, $D_{\min} \geq 1$ to the lower limit of fractal dimension, $p$ is a scaling parameter associated with the initial growth rate $k$, and $t_0$, a temporal translational parameter indicative of a critical time, when the rate of fractal dimension growth indicating city growth reaches its peak. The scale and scaling parameters can be respectively defined by

$$p = \frac{1}{k}, \quad t_0 = \ln[\frac{D_{\max} - D_{(0)}}{D_{(0)} - D_{\min}}]^p.$$

For the normalized data, equation (1) can be re-expressed as

$$D^*(t) = \frac{D(t) - D_{\min}}{D_{\max} - D_{\min}} = \frac{1}{1+(1/D^*_{(0)} - 1)e^{-kt}}, \quad (2)$$

where $D^*_{(0)} = (D_{(0)} - D_{\min})/(D_{\max} - D_{\min})$ denotes the normalized result of $D_{(0)}$, the original value of fractal dimension. This is a logistic model.

Empirically, equations (1) and (2) can be supported and thus validated by the dataset of London from Batty and Longley (1994), and by the datasets of Tel Aviv from Benguigui *et al* (2000) as well as the dataset of Hangzhou from Feng and Chen (2010) (Appendix 2). The derivative of equation (2) is just the logistic equation

$$\frac{dD^*(t)}{dt} = kD^*(t)[1-D^*(t)]. \quad (3)$$

For simplicity, and without loss of generality, let the time interval $\Delta t = 1$. Thus, discretizing equation (3) yields a 1-dimensional map such as

$$D^*_{t+1} = (1+k)D^*_t - kD^{*2}_t. \quad (4)$$

Defining $D^*_t = (1+k)x_t/k$, we can transform equation (4) into the following form

$$x_{t+1} = (1+k)x_t(1-x_t) = \mu x_t(1-x_t). \quad (5)$$

where $x_t$ is the substitute of $D^*_t$, and $\mu = k+1$ is a growth rate parameter. Equation (5) is just the well-known logistic map, which can generate complex dynamics such as periodic oscillations and chaos (May, 1976). If the fractal dimension of urban form can be fitted to Boltzmann's equation, it will imply that urban evolution can be associated with spatial chaotic dynamics.

**2.2 Level of urban space-filling of cities**

Fractal dimension can employed to define various urban form indexes indicating urban shape, size and development (see e.g. Chen, 2011; Chen and Lin, 2009; Longley *et al*, 1991; Thomas *et al*, 2010). Next, let's define a spatial replacement index by using the normalized fractal dimension. The logistic function of fractal dimension implies an urban spatial substitution equation, which suggests the process that the natural/rural land is replaced by urban built-up area. From equation (2) follows that

$$\ln[\frac{D^*(t)}{1-D^*(t)}] = \ln(\frac{D_0^*}{1-D_0^*}) + kt, \tag{6}$$

where $0<D^*<1$. We can define a spatial *filled-unfilled ratio* (FUR) for urban growth, that is

$$O = \frac{D^*}{1-D^*} = \frac{U}{V}. \tag{7}$$

Thus we have

$$D^* = \frac{O}{O+1} = \frac{U}{U+V} = \frac{U}{S}. \tag{8}$$

where $U$ refers to the filled space area with various urban buildings (space-filling area), measured by the pixel number of urbanized land on digital maps, and $V$, to the unfilled space area without any construction or artificial structures (space-saving area). Thus the total space of urbanized region is $S=U+V$. Obviously, the higher the $O$ value is, the higher the degree of urban spatial filling is. The normalized fractal dimension can be termed *level of space filling* (LSF) of cities, implying the degree of spatial replacement.

If we make measurements on digital maps, the filled space area, $U$, will be a known number. However, the unfilled space area, $V$, may be unknown. On the other, if we can estimate $D$ value and thus $D^*$ value, then we can estimate the $V$ values and thus $S$ value (Appendix 3). According to equation (8), the unfilled space area can be given by

$$V = U(\frac{1}{D^*} - 1). \tag{9}$$

Consequently, the total spatial area of a city can be given by

$$S = U + V = \frac{U}{D^*}. \tag{10}$$

The above-defined urban spatial measurements and related formula are useful for urban analysis in practice.

**2.3 From 1-D map to 2-D map**

The process of urban growth is a dynamic process of urban space filling. The natural space or rural land is gradually replaced by artificial space or urban land. That is, the vacant space (e.g., open space, spare space) in a city is continually filled in by various urban infrastructure and superstructure. If a region is extensively developed and is already occupied by urban structures, outbuildings, and service areas, it is urbanized, and the rural space is replaced by urban space. This spatial replacement dynamics can be described by a pair of differential equations

$$\frac{dU(t)}{dt} = \alpha U(t) + \beta \frac{U(t)V(t)}{U(t) + V(t)}, \tag{11}$$

$$\frac{dV(t)}{dt} = \lambda V(t) - \beta \frac{U(t)V(t)}{U(t) + V(t)}, \tag{12}$$

where $\alpha$, $\beta$, and $\lambda$ are parameters. This implies that the growth rate of filled space, $dU(t)/dt$, is proportional to the size of filled space, $U(t)$, and the coupling between filled and unfilled space, but not directly related to unfilled space size; the growth rate of unfilled space, $dV(t)/dt$, is proportional to the size of unfilled space, $V(t)$, and the coupling between unfilled and filled space, but not directly related to filled space size. From equations (11) and (12) follows equation (3). Actually, taking derivative of equation (8) yields

$$\frac{dD^*(t)}{dt} = \frac{dU(t)/dt}{U(t) + V(t)} - \frac{U(t)}{[U(t) + V(t)]^2}\left[\frac{dU(t)}{dt} + \frac{dV(t)}{dt}\right]. \tag{13}$$

Substituting equations (11) and (12) into equation (13) gives

$$\frac{dD^*(t)}{dt} = \frac{\beta U(t)V(t)}{[U(t) + V(t)]^2} + \frac{U(t)}{U(t) + V(t)}(\alpha - \frac{\alpha U(t)}{U(t) + V(t)} - \frac{\lambda V(t)}{U(t) + V(t)}). \tag{14}$$

Further consider equation (8), we have

$$\frac{dD^*(t)}{dt} = (\alpha + \beta - \lambda)D^*(t)[1 - D^*(t)]. \tag{15}$$

Comparing equations (13) with equation (3) shows

$$k = \alpha + \beta - \lambda. \tag{16}$$

This suggests that equations (11) and (12) are equivalent to equation (3). That is, the dynamical analysis based on the logistic equation can be displaced by the dynamical analysis based on the space replacement model. In fact, equations (11) and (12) can be discretized and become a 2-dimensional map such as

$$U_{t+1} = (1 + \alpha)U_t + \beta \frac{U_t V_t}{U_t + V_t}, \tag{17}$$

$$V_{t+1} = (1 + \lambda)V_t - \beta \frac{U_t V_t}{U_t + V_t}, \tag{18}$$

where $U_t$ and $V_t$ are the discrete expression of $U(t)$ and $V(t)$, respectively. This suggests that a 1-dimensional map can be converted into a 2-dimensional map.

## 3. Method and results

### 3.1 The 1-D logistic map

The logistic map is in fact a polynomial mapping of degree 2. The map was popularized in a seminal paper of May (1976), and it is usually cited as an archetypal example of how complicated behaviors such as oscillation and chaos can arise from very simple non-linear dynamical equations. Equation (3) is identical in form to the discrete-time demographic model which was first created by Pierre François Verhulst (Banks, 1994). The recurrence relation, equation (5), is just the well-known logistic map. For comparison with the results in the following subsection, the dynamical behaviors from fixed state to periodic oscillation, and then to chaos are illustrated in Figure 1, which shows the logistic map for 60 generations of $D^*(t)$ as $k$ moves from 1.65 to 2.95. The mapping is so simple that we can carry out the numerical calculation by means of MS Excel.

It is not novel for us to see Figure 1, but it is interesting to compare Figure 1 with Figure 2. Figure 1 is based on the 1-dimensional logistic map, while Figure 2 is based on the 2-dimensional interaction map. As indicated above, the 1-dimensional map, equation (4), can be derived from the 2-dimensional maps, equations (17) and equation (18). The similarities between Figure 1 and Figure 2 suggest the inherent relation between the 1-dimensional and the 2-dimensional map, whereas the differences between them denote that the 2-dimensional map says more than the 1-dimensional map. Actually, the spatial replacement maps can exhibit more simple and complex behaviors than those arising from the logistic map.

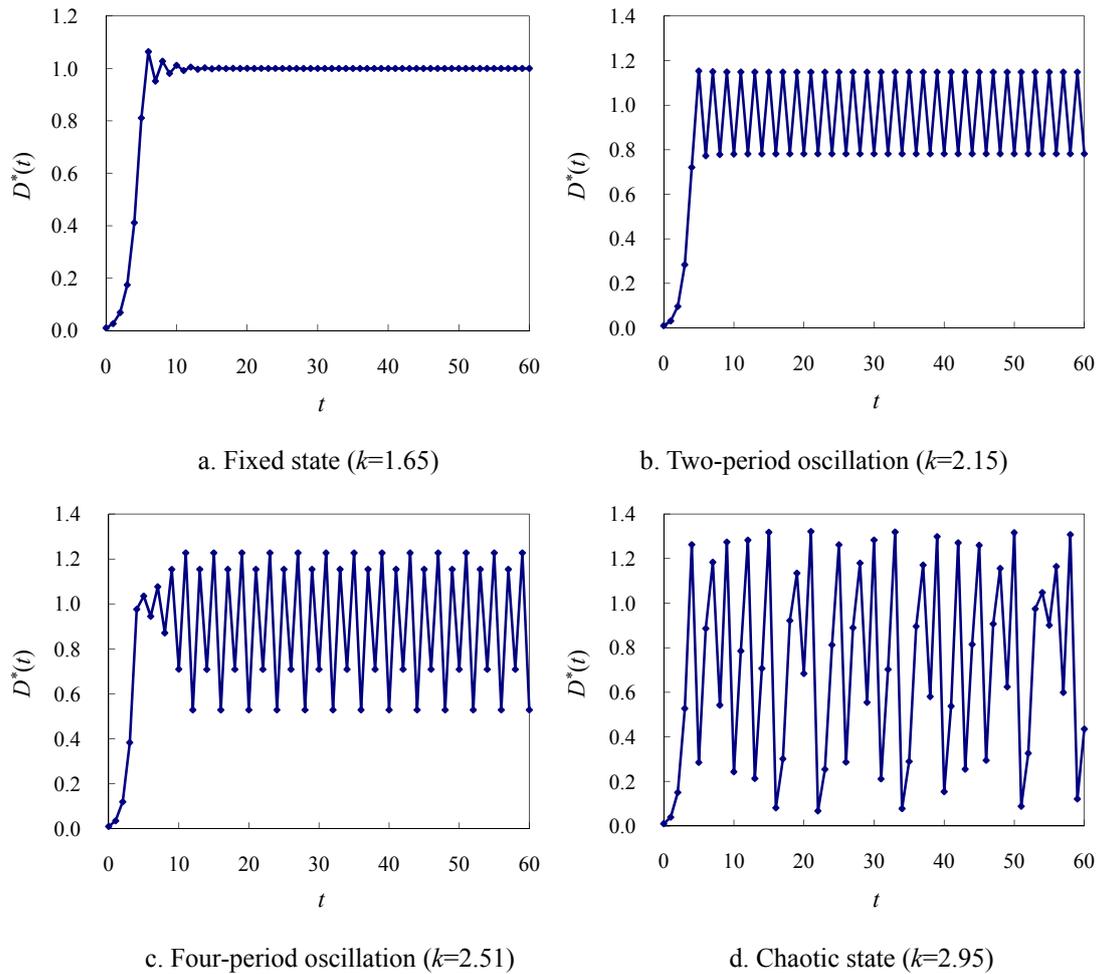

a. Fixed state ($k=1.65$)　　　　　　b. Two-period oscillation ($k=2.15$)

c. Four-period oscillation ($k=2.51$)　　　　　　d. Chaotic state ($k=2.95$)

**Figure 1 The fractal dimension growth of urban form based on the 1-dimensional map: from fixed state to oscillation, and then to chaos** (This is the same patterns as that from May (1976))

### 3.2 The 2-D space-replacement map

Among the three parameters of the space replacement maps, $\alpha$, $\beta$, and $\lambda$, the second one, $\beta$, is the most important because it controls the rate of space filling and replacement. For simplicity, we can fix the parameter value of $\alpha$, and $\lambda$, for instance, let $\alpha=0.015$, $\lambda=0.005$. Then, if we change $\beta$ value, $U_t$ value and $V_t$ value will change, and accordingly, $D_t^*$ value will change. In the process of changing $\beta$ value, various simple and complex behaviors appear, including S-shaped growth, two-period oscillation, four-period oscillation,…, and chaos (Figure 2). Comparing Figure 1 with Figure 2 shows that the 1-dimeanional map reflects only a special case of the 2-dimensional map. The 2-dimensional non-linear dynamics is richer and more colorful than the 1-dimensional non-linear dynamics.

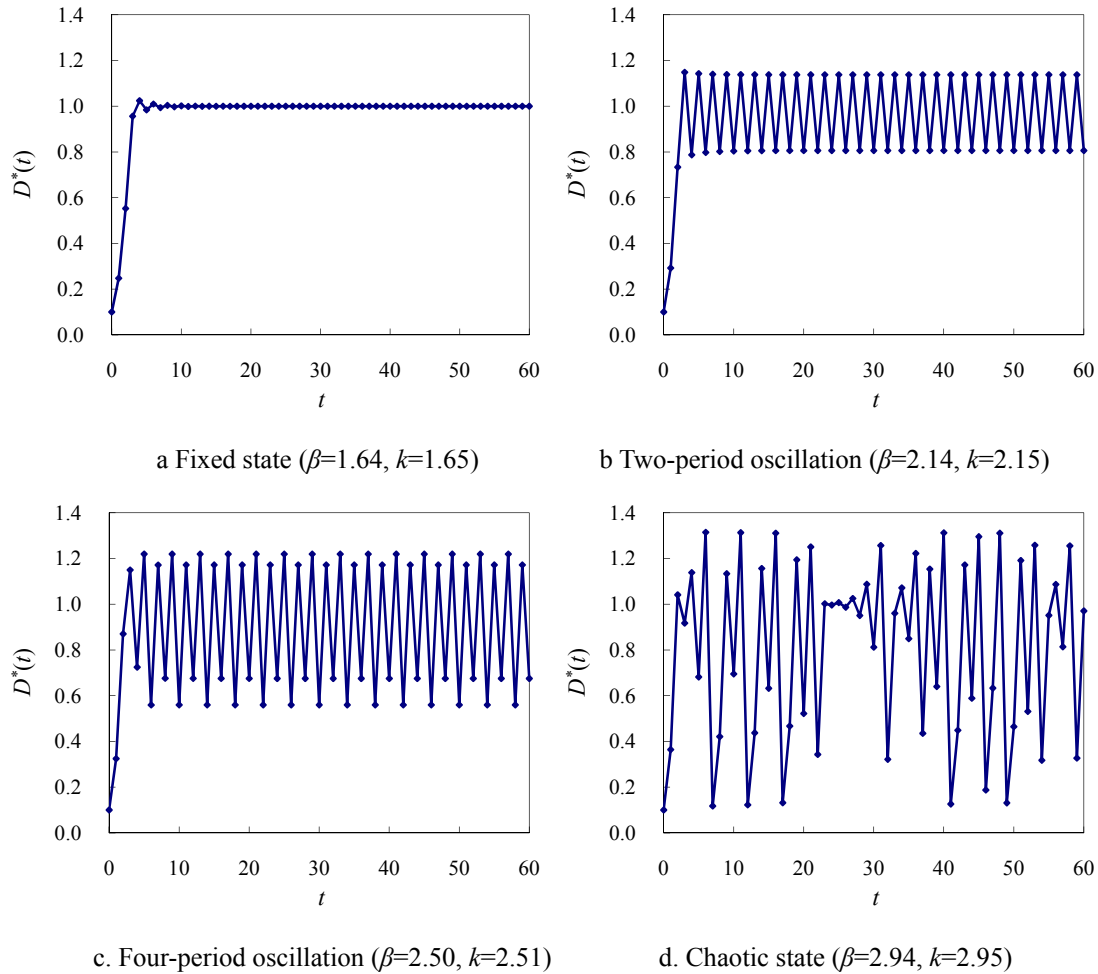

a Fixed state ($\beta$=1.64, $k$=1.65)   b Two-period oscillation ($\beta$=2.14, $k$=2.15)

c. Four-period oscillation ($\beta$=2.50, $k$=2.51)   d. Chaotic state ($\beta$=2.94, $k$=2.95)

**Figure 2 The fractal dimension growth of urban form based on the 2-dimensional map: from fixed state to oscillation, and then to chaos** (this is the same patterns as that from Chen (2009))

The course form period doubling bifurcation to chaos can be shown in the phase space based on $U(t)$ vs $V(t)$. First, a curve, corresponding to sigmoid growth or fixed state; then two straight lines, corresponding to two-period oscillation; then four straight lines, corresponding to four-period oscillation, and so on; and finally, numerous scattered points or arbitrary number of straight lines suggesting chaos (Figure 3). The chaotic state is always reflected by two kinds of phase portraits.

One is the random data points which are confined by two cross straight lines, and the other, the straight lines with number not equal to $2^n$ ($n$=1, 2, 3, …), say 3 lines, 5 line, 6 lines. This implies that there exists no strange attractor for this chaos.

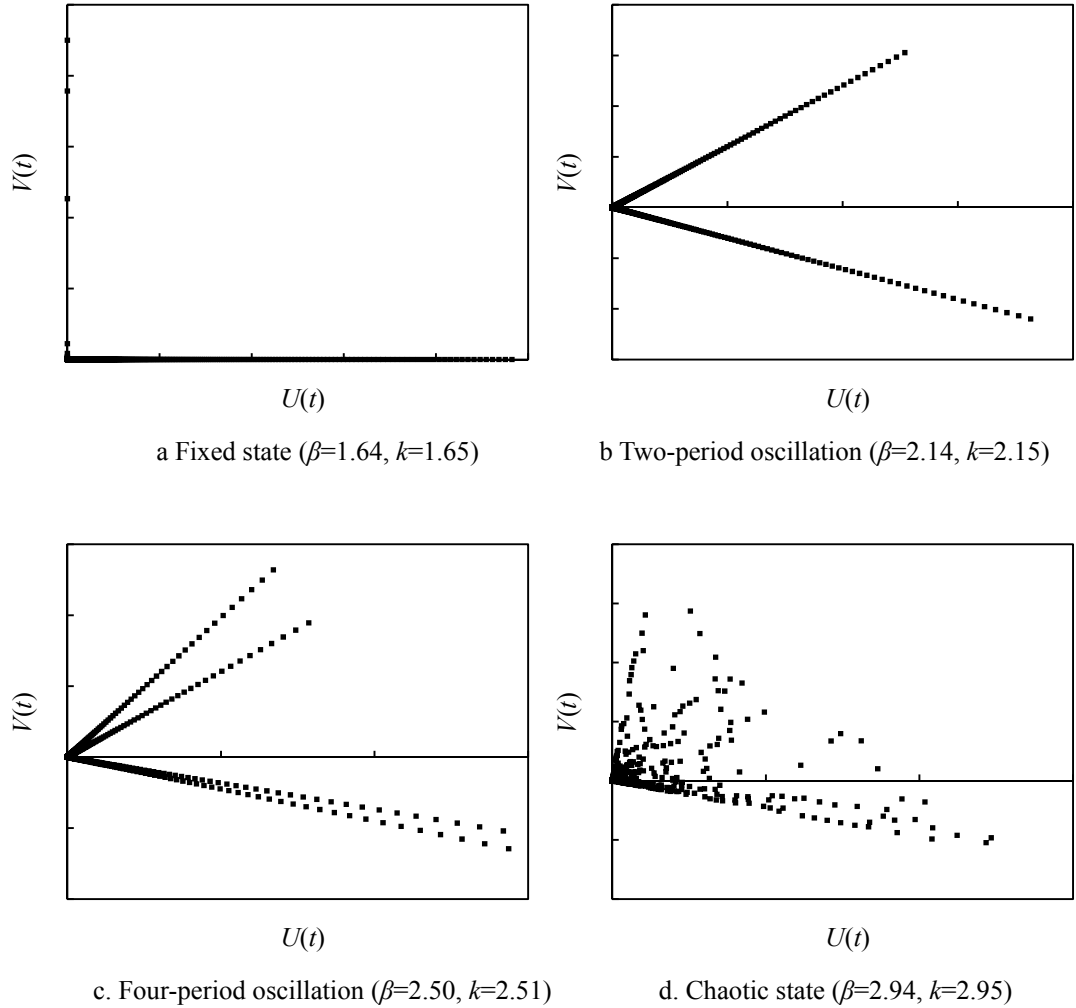

a Fixed state ($\beta$=1.64, $k$=1.65)  
b Two-period oscillation ($\beta$=2.14, $k$=2.15)  
c. Four-period oscillation ($\beta$=2.50, $k$=2.51)  
d. Chaotic state ($\beta$=2.94, $k$=2.95)

**Figure 3 Four patterns of phase space corresponding the fixed state, oscillation, and chaos (1500 data points)**

## 4. Discussion

A new way of looking at the relation between fractals and chaos is presented here. The fractal dimension growth of growing fractals may be associated with chaotic dynamics. A typical regular growing fractal is displayed in Figure 4, which was proposed by Jullien and Botet (1987) and Vicsek (1989) and employed to model urban growth in theory by Batty and Longley (1994), Frankhauser (1998), and Longley et al (1991). The typical simple random growing fractal is the diffusion-limited aggregation (DLA) cluster shown in Figure 5, which is employed to simulate urban growth in practice by Batty *et al* (1989) and Fotheringham *et al* (1989). Though urban form can be modeled with fractal geometry, and DLA is metaphor of a city, a growing fractal is usually different from urban growth (Table 1). For the growing fractal in mathematics, the fractal dimension never change over time represented by steps. However, for urban growth, the fractal dimension is function of time. As indicated above, the fractal dimension evolution can be described by Boltzmann's equation.

Table 1 Comparison between city fractals and mathematical fractals

| Item | Mathematical fractal | City fractal |
| --- | --- | --- |
| **Fractal** | Simple structure | Complex structure |
| **Fractal dimension** | Invariable--Constant of time | Variable—function of time |
| **Pattern** | Regular or irregular pattern | Irregular pattern |
| **process** | Certain or random process | Random process |

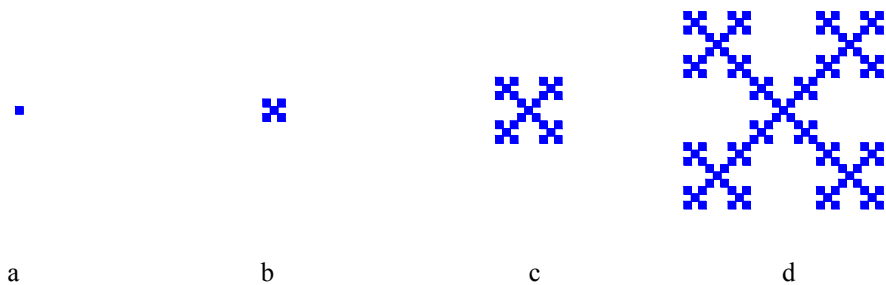

a            b            c            d

**Figure 4 A regular growing fractals which are often employed as model of urban growth**

This paper is devoted to reveal the essence of the fractal dimension of urban form and its evolution regularity. Especially, this work is also devoted to developing a theory of urban spatial replacement. The mains of this theory are as follows. Urban space falls into two parts. One part is the filled space, and the other, the unfilled space (Table 2). The process of urban growth is in fact a process of logistic spatial substitution. That is, the replacement of natural space by artificial space, or of rural land by built-up land. The normalized fractal dimension of urban form is in fact the ratio of the filled space to the total urban space, and can be described with the logistic function.

The rate of spatial replacement should be controlled and kept in an appropriate scale. If the spatial replacement rate exceed some critical or threshold value, the periodic oscillations or even chaos will arise suddenly, and the city system will lose its balance or even fall into confusion. The suggestions of city planning and policy can be made as below. The proper proportion between filled space and unfilled space should be planned for urban space, and the rate of urban growth should be controlled for keeping spatial order in a city.

Table 2 The filled space and unfilled space of a city and the contents

| Item | Filled space | Unfilled space |
| --- | --- | --- |
| **Space** | Urbanized area, built-up area, buffer space, etc. | Vacant space, spare space, open space, etc. |
| **Land** | Industrial land, commercial land, transport land, residential land, storage land, etc. | Farmland, woodland |
| **Objects** | Structures, outbuildings, service areas, etc. | Greenbelt, water body, etc. |
| **Set** | Fractal set | Fractal complementary set |

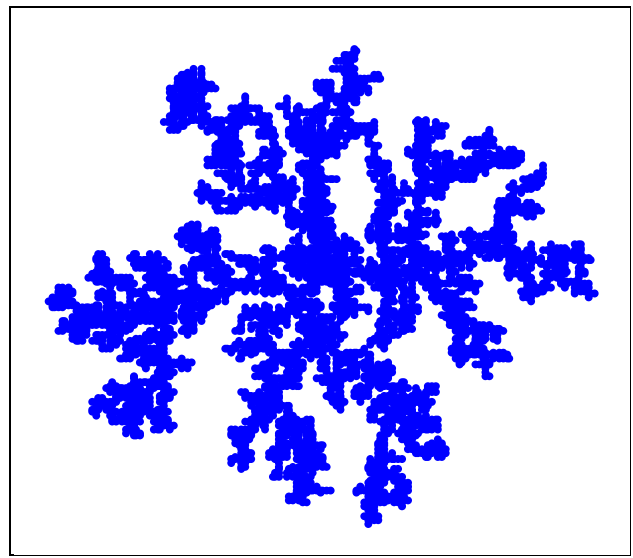

Figure 5 A random growing fractal which is often used to simulate urban growth—A DLA cluster with 5000 particles (generated by Matlab)

# 5. Conclusions

The theoretical background and empirical base of this paper is that the fractal dimension growth of urban form can be modeled with Boltzmann's equation, which is equivalent to the logistic function for the normalized data. We can fit the fractal dimension dataset of London from Batty and Longley (1994), the dimension datasets of Tel Aviv from Benguigui *et al* (2000), and the datasets of Hangzhou from Feng and Chen (2010) to the Boltzmann's equation. The well-known logistic equation suggests spatial replacement dynamics for urban growth.

By using mathematical method, we can derive a 1-dimenaional logistic map from the Boltzmann's equation. On the other, from the 2-dimensional map indicative of a spatial replacement process follows the 1-dimensional logistic map. A finding is that the fractal dimension of urban form can be treated as a ratio of space filling of a city. By means of numerical computation, we can reveal that the rate of spatial replacement dominate the nonlinear dynamics of fractal dimension evolution. The fractal dimension growth and chaotic behavior can be linked by spatial replacement dynamics.

A conclusion can be drawn that the urban space can be divided into two parts--filled space and unfilled space. The normalized fractal dimension of urban form is just the ratio of the filled space to the total urban space. If the rate of urban growth is too high, the periodic oscillations or chaos will emerge and thus urban evolution will fall into disorder. The significance of this study is as follows. First, it provides a new way of looking at the relation between chaos and fractals. Second, it provides a new angle of view of making urban spatial analysis. Third, it provides a case to support the spatial replacement dynamics. The urban space-filling dynamics can be extended to a general theory of spatial replacement dynamics, which can be employed to explain the spatio-temporal evolution of many complex systems.

## Acknowledgements

This research was sponsored by the Natural Science Foundation of Beijing (Grant No. 8093033) and the National Natural Science Foundation of China (Grant No. 40771061).

# Appendices

## A1 Squashing effect and sigmoid function

If some kind of measure of a system has clear upper limit and lower limit, and if growth of the system is not of uniform speed, the growing course of the measure always takes on the S-shaped curve. Formally, the curve can be abstracted as what is called *sigmoid function*. Sometimes, the sigmoid function is called *squashing function*. Pressed by the upper limit and withstood by the lower limit, a line will be twisted into S shape. Thus a straight line will change to a sigmoid curve. The family of sigmoid functions includes various functions such as the ordinary arc-tangent, the hyperbolic tangent, and the error function, the Gompertz function, and the generalized logistic function (say, Boltzmann's equation, quadratic logistic function). Among various sigmoid functions, the logistic function is very familiar to scientists and engineers. The level of urbanization ranging from 0 to 1 can often be fitted to the logistic function since it has clear upper and lower limits. Similarly, the fractal dimension of urban form based on 2-dimensional digital maps always ranges from 1 to 2. In other words, the fractal dimension has clear upper limit ($D_{max}=2$) and lower limit ($D_{min}=1$). So, it is possible that fractal dimension of urban form follows the law of logistic growth, and can be fitted to Boltzmann's equation.

## A2 Two examples of fitting real datasets to Boltzmann's equation

The examples of London and Tel Aviv can be employed to testify equation (1). The fractal dimension evolution of the London can be modeled with Boltzmann's equation. The fractal dimension dataset of urban form comes from Batty and Longley (1994), who provided the fractal dimension values of the largest city of the United Kingdom (UK) from 1820 to 1962. For simplicity, let $D_{min}=1$, then a least square computation gives a model for London such as

$$D(t) = 1 + \frac{0.806}{1+\exp(-\frac{t-5.302}{34.266})}.$$

The goodness of fit is about $R^2=0.847$ (Figure A1a).

The second case is on the urban form of Tel Aviv, the secondary most populous city in Israel. The dataset of fractal dimension comes from Benguigui et al (2000). In order to determine when and where a city fractal is, Benguigui et al (2000) defined three study regions on the digital maps of Tel Aviv metropolis and estimated the fractal dimension from 1935 to 1991 by using the box-counting method. For example, for Region 2, the model is

$$D(t) = 1 + \frac{0.845}{1+\exp(-\frac{t-4.148}{27.993})}.$$

The correlation coefficient square is around $R^2=0.990$ (Figure A1b). The model can also be applied to Hangzhou, a well-known China's city. The datasets came from Feng and Chen (2010).

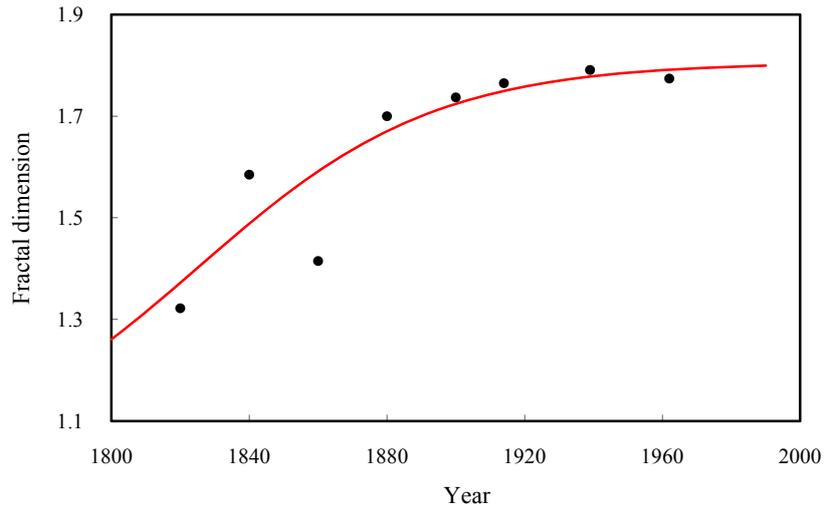

a. London, 1820-1962

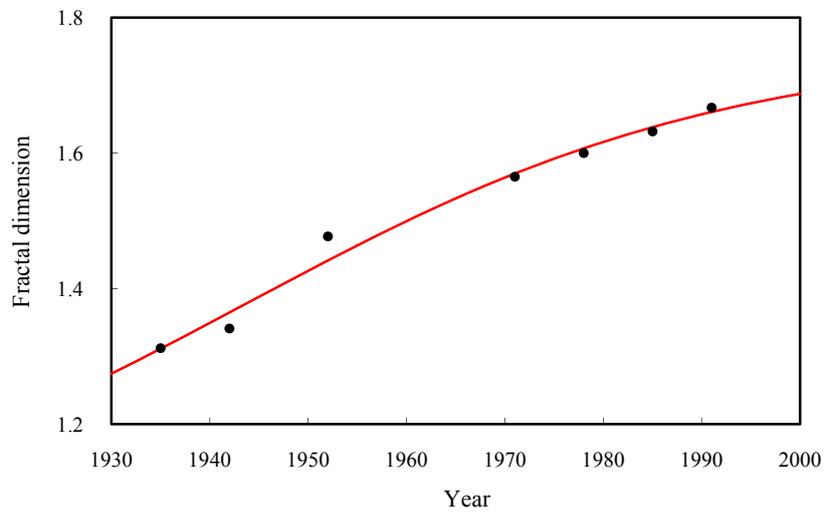

b. Tel Aviv, 1935-1991

**Figure A1 The fractal dimension growth of London and Tel Aviv modeled by Boltzmann's equation**

**A3 Spatial replacement index and urban space classification**

Applying the formula of urban spatial replacement index, equation (7) and (8), to the national capital of China, Beijing, yields the results as follows (Table A1).

**Table A1 The filled area, unfilled area, fractal dimension, and related measurements of Beijing**

| Year | Filled area ($U$) | Unfilled area ($V$) | Total space ($S$) | Box dimension ($D$) | LSF ($D^*$) | FUR ($O$) |
|---|---|---|---|---|---|---|
| 1988 | 837158000 | 146909937 | 984067937 | 1.851 | 0.851 | 5.698 |
| 1992 | 851491010 | 140488037 | 991979047 | 1.858 | 0.858 | 6.061 |

| 1999 | 1140490000 | 127065981 | 1267555981 | 1.900 | 0.900 | 8.976 |
| 2006 | 1536300000 | 116176417 | 1652476417 | 1.930 | 0.930 | 13.224 |
| 2009 | 1907170000 | 84715122 | 1991885122 | 1.957 | 0.957 | 22.513 |

**Note**: The study area is $A$=3331624752, $D_{max}$=2, $D_{min}$=1. The $U$ values are measured on digital maps, while the $V$ values are estimated by $U$ and the fractal dimension $D^*$ values. The $O$ values are equivalent to the $D^*$ values.